\documentclass[graybox]{svmult}
\usepackage[utf8]{inputenc} 

\usepackage{type1cm}        
\usepackage{makeidx}         
\usepackage{graphicx}        
                             
\usepackage{multicol}        
\usepackage{multirow}
\usepackage[bottom]{footmisc}

\usepackage{newtxtext}       %
\usepackage{newtxmath}       

\makeindex             

\begin{document}

\graphicspath{{figures/}}

\title*{Fluid-structure interaction using volume penalization and mass-spring models with application to flapping bumblebee flight}
\titlerunning{Fluid-structure interaction: application to bumblebee flight}
\author{Hung Truong, Thomas Engels, Dmitry Kolomenskiy and Kai Schneider}
\institute{Hung Truong \at Aix-Marseille Universit\'e, CNRS, Centrale Marseille, I2M, Marseille, France\\ \email{dinh-hung.truong@univ-amu.fr}
\and Thomas Engels \at LMD-IPSL, Ecole Normale Sup\'erieure-PSL, Paris, France\\ \email{thomas.engels@ens.fr}
\and Dmitry Kolomenskiy \at Japan Agency for Marine-Earth Science and Technology (JAMSTEC), Yokohama, Japan\\ \email{dkolomenskiy@jamstec.go.jp}
\and Kai Schneider \at Aix-Marseille Universit\'e, Centrale Marseille, I2M, Marseille, France\\ \email{kai.schneider@univ-amu.fr}}
%
%
\maketitle

\abstract{Wing flexibility plays an essential role in the aerodynamic performance of insects due to the considerable deformation of their wings during flight under the impact of inertial and aerodynamic forces. These forces come from the complex wing kinematics of insects. In this study, both wing structural dynamics and flapping wing motion are taken into account to investigate the effect of wing deformation on the aerodynamic efficiency of a bumblebee in tethered flight. A fluid–structure interaction solver, coupling a mass–spring model for the flexible wing with a pseudo-spectral code solving the incompressible Navier--Stokes equations, is implemented for this purpose. We first consider a tethered bumblebee flying in laminar flow with flexible wings. Compared to the rigid model, flexible wings generate smaller aerodynamic forces but require much less power. Finally, the bumblebee model is put into a turbulent flow to investigate its influence on the force production of flexible wings.}

\section{Introduction}
\label{sec:intro}

In recent years, the effect of wing flexibility on aerodynamic performance of flapping wings has drawn attention of researchers, scientists and engineers. Compared to conventional airplanes with fixed wings, flapping wings have several aerodynamic advantages with the ability to create lift even at high angles of attack due to the delayed stall of the leading edge vortex (LEV)~\cite{EllingtonLEV}. These extraordinary capabilities of hovering and manoeuvrability have made bio-inspired flapping wings a strong candidate for developing human-engineered micro-air-vehicles (MAVs) with possible applications in environmental monitoring, surveillance and security. However, in previous studies, the wings were usually considered as rigid to simplify the problem. The sophisticated interaction between the anisotropic wing structures and the surrounding unsteady flow makes the analysis of flapping flexible wings challenging, but at the same time intriguing. In the last two decades, with the dramatic improvement of measuring equipment as well as computing power, many experimental and numerical studies have investigated the effect of wing flexibility and drawn contradictory conclusions. Mountcastle and Combes~\cite{Combesflexbb} showed  that  passive  deformations enhance lift production in bumblebees by artificially stiffening their wings using a micro-splint. Campos et al.~\cite{CamposFlexFlatPIV} and Fu et al.~\cite{WShyyFlexAeroPeform} used experimental methods and found that highly flexible wings show significant tip-root lag which weakened vortices and reduce the force production. Du and Sun~\cite{Du} solved the Navier--Stokes equations coupled with measured wing deformation data and compared with the rigid counterparts. They obtained a $10 \%$ increase in lift caused by the camber deformation and a $5 \%$ reduction in required power.

In our previous work~\cite{HungRevol}, we considered a flexible bumblebee wing rotating around a hinge point at the angle of attack equal to $45^{\circ}$. The stiffness of the wing was varied to get two cases: flexible and highly flexible. We found that the flexible wing produces less lift than the rigid wing but it has a better lift-to-drag ratio. On the other hand, the highly flexible wing experienced a strong tip-root lag caused by twisting and behaves poorly in term of aerodynamic performance with a much smaller lift and lift-to-drag ratio. Although the study provided us some ideas about the influence of wing flexibility on the force generation, the revolving motion remains too simple to fully represent the complicated dynamics of a flapping wing. After a short transition period, the revolving wing attains its steady state and its dynamic deformation hardly plays a role in the force production. The wing kinematics of insects is in reality more intricate with many characteristic features such as flapping amplitude, wingbeat frequency, angle of attack, etc. These features have strong impact on the ability of generating force of the wings. Kang and Shyy~\cite{WShyyLiftFlex} showed that the ratio between the flapping frequency and the first natural frequency of a flexible wing can yield advanced, symmetric or delayed rotation modes which in turn alter the resulting lift. Zhao et al.~\cite{ZhaoAeroFlexFlap} conducted experiments of simple isotropic flapping wings with varied stiffness values at different angles of attack. They found that at low angles of attack ($20^{\circ}$ to $60^{\circ}$), flexible wings have relatively the same aerodynamic performance as rigid wings but they outperformed their rigid counterparts at high angles of attack (up to $90^{\circ}$).

Consequently, in this work, we will investigate the aerodynamic efficiency of the flexible bumblebee wing model~\cite{HungRevol} within a tethered flight context where the wing motion is real bumblebee wing kinematics measured by the work of Dudley and Ellington~\cite{DudleyBBWngKinematics}. The resulting force and power are compared with those of a rigid flat wing computed by Engels et al.~\cite{EKSLS16}. The comparison shows the effects of the wing deformation on the aerodynamic forces of a flapping flight insect.

The remainder of the manuscript presents in section~\ref{sec:nummeth} the numerical methods used for solving the governing equations of the flexible wing, the fluid flow and its coupling.
The numerical set-up and the bumblebee model are given in section~\ref{sec:setupbb} and the results of the numerical simulations are discussed in section~\ref{sec:results}.
Finally some conclusions are drawn in section~\ref{sec:conclusions}.

\section{Numerical methods and governing equations}
\label{sec:nummeth}

Modeling insect flight is a delicate topic due to the fact that one needs to model both the mechanical behavior of the wings by a solid solver and the surrounding flow by a fluid solver. The two solvers must then be coupled to study their interaction. This section presents how these three aspects can be handled numerically.

\subsection{Solid solver using mass-spring system}

 Insect wings are sophisticated structures consisting of membranes and veins. The wings get their nonlinear anisotropic properties from a truss framework composed of horizontal and vertical veins connected by membranes~\cite{ShyyIntroFlap}. This along with their small wing lengths (from $mm$ to $cm$) makes it extremely challenging to model the mechanical behavior of insect wings. In our work, a mass-spring system is employed to mimic the dynamics of the complicated membrane-vein network by taking  the different mechanical properties between veins and membranes into account~\cite{HungRevol}. While veins can be considered as rods which resist mainly the torsion and bending deformation, a membrane is fabric-like and behaves like a piece of cloth which resists against the extension deformation.

 The mass-spring system has been around since the end of the $20^{th}$ century and it is well-known for its  computational efficiency and ability of handling large deformation~\cite{DeforModelNealen}. The wing is discretized using mass points $m_i \, (i=1,...,n)$ connected by springs. Among the different types of springs, our model is built only on extension and bending springs. The dynamic behavior of the mass-spring system, at a given time $t$, is defined by the position $\mathbf{x}_i$ and the velocity $\mathbf{v}_i$ of the mass point $i$ and they are governed by the eqns. (\ref{eqn:Newton_law}) below:

\begin{equation}
\begin{split}
\mathbf{F}^{\mathrm{int}}_i + \mathbf{F}^{\mathrm{ext}}_i & = m_i \mathbf{a}_i  \ \ \ \ \ \ \rm{for} \ \ \mathit{i = 1 \ldots n}  \\
\mathbf{v}_i (t=0) & = \mathbf{v}_{0,i}\\
\mathbf{x}_i (t=0) & = \mathbf{x}_{0,i}
\end{split}
\label{eqn:Newton_law}
\end{equation}
where $n$ is the number of mass points, $\mathbf{F}^{\mathrm{int}}_i$ is the internal force and $\mathbf{F}^{\mathrm{ext}}_i$ is the external force acting on the $i^{th}$ mass point, $m_i$ and $\mathbf{a}_i$ are mass and acceleration of the $i^{th}$ mass point, respectively.

The system (\ref{eqn:Newton_law}) is then advanced numerically in time by applying a second order backward differentiation scheme with variable time steps \cite{BDFscheme}:

\begin{equation}
\begin{split}
\mathbf{q}_i^{n+1} - \frac{(1+\xi)^2}{1+2\xi} \mathbf{q}_i^{n} + \frac{\xi^2}{1+2\xi} \mathbf{q}_i^{n-1} & = \frac{1+\xi}{1+2\xi} \Delta t^n \mathbf{f}(\mathbf{q}_i^{n+1}) \\
\end{split}
\label{eqn:discretized_ODEs}
\end{equation}

where $\mathbf{q} = \big[ \mathbf{x}_i, \ \ \mathbf{v}_i \big]^\intercal$ is the phase vector containing positions and velocities of all mass points and $\mathbf{f} (\mathbf{q}) = \big[ \mathbf{v}_i, \ \ m_i^{-1} (\mathbf{F}^{\mathrm{int}}_i + \mathbf{F}^{\mathrm{ext}}_i) \big]^\intercal$ is the right hand side function, $\xi=\Delta t^n / \Delta t^{n-1}$ is the ratio between the current time step $\Delta t^n$ and the previous one $\Delta t^{n-1}$. The phase vector of the system at the current time step $\mathbf{q}^{n+1}$ is found by solving eqn. (\ref{eqn:discretized_ODEs}) using the Newton--Raphson method. All details of this solver are explained in \cite{HungRevol}.

\subsection{Fluid solver and volume penalization method}

Due to their small sizes and elevated flapping frequencies, insect flight is normally categorized in the Reynolds number regime between $\mathcal{O}(10^1)$ and $\mathcal{O}(10^4)$. For example, for hawkmoth we have $Re=6000$, bumblebee $Re=2000$, fruit fly $Re=100$ or thrips $Re=10$~\cite{Shyy2016,EKSFLS19}. The flow can be considered as incompressible and governed by: 
\begin{align}
\partial_t \mathbf{u} + \text{\boldmath$\omega$} \times \mathbf{u} &= - \nabla \Pi + \nu \nabla^2 \mathbf{u} - \underbrace{\frac{\chi}{C_{\eta}} (\mathbf{u} - \mathbf{u}_s)}_{\text{penalization term}} - \underbrace{\frac{1}{C_{\mathrm{sp}}} \nabla \times \frac{(\chi_{sp} \text{\boldmath$\omega$})}{\nabla^2}}_{\text{sponge term}} \label{eqn:penalized_NS_momentum} \\
\nabla \cdot \mathbf{u} &= 0 \label{eqn:penalized_NS_continuity} \\
\mathbf{u} (\mathbf{x}, t=0) &= \mathbf{u}_0 (\mathbf{x}) \ \ \ \ \ \ \mathbf{x} \in \Omega , \, t>0
\label{eqn:penalized_NS_ini}
\end{align}
The above equations (\ref{eqn:penalized_NS_momentum}-\ref{eqn:penalized_NS_ini}) are called the penalized Navier--Stokes equations~\cite{VolPenaAngot} where $\mathbf{u}$ is the fluid velocity, $\text{\boldmath$\omega$} = \nabla \times \mathbf{u}$ is the vorticity, $\Pi = p +\frac{1}{2}\mathbf{u} \cdot \mathbf{u}$ is the total pressure and $\nu$ is the kinematic viscosity. Except for all the terms found in the classical incompressible Navier--Stokes equations, it appears two more terms which are called the sponge and the penalization terms. The former is added to remove the periodicity of the Fourier discretization which affects the upstream inflow. The penalization term is used to impose the no-slip boundary conditions on the fluid-solid interface in~\cite{Flusi}. All geometrical information of the solid is encoded in the mask function $\chi$ given by:
\begin{equation}
    \chi(\delta)= 
\begin{cases}
    1 &  \delta \leq d-h\\
    \frac{1}{2} \left(1+\cos{\pi \frac{(\delta-d+h)}{2h}}\right) & d-h < \delta < d+h \\
    0 & \delta \geq d+h
\end{cases}
\label{eqn:mask_function}
\end{equation}
where $\delta$ is the signed distance field of the bumblebee skeleton and $d$ represents here the distance from the skeleton to the outer surface, i.e. the fluid-solid interface. The skeleton of the bumblebee is a curvilinear centerline along which we sweep an elliptical section of variable size to draw the insect's body, legs and antennae~\cite{ThomasThesis}. However, to avoid the force oscillation when dealing with moving solid body, a smoothing layer with a thickness $2h$ is added right at the fluid-solid interface to prevent the discontinuity of the mask function~\cite{SmoothMask}.

For solving the fluid equations (\ref{eqn:penalized_NS_momentum}-\ref{eqn:penalized_NS_ini}), a Fourier pseudospectral discretization with semi-implicit time stepping is employed, implemented in the FLUSI \footnote{FLUSI: freely available for noncommercial use from GitHub (https://github.com/pseudospectators/FLUSI).} code \cite{Flusi}. The general idea consists of representing quantities $q$ (velocity, pressure, vorticity) as truncated Fourier series,

\begin{equation}
    q(\mathbf{x},t) = \sum_{k_x=0}^{N_x-1}  \sum_{k_y=0}^{N_y-1} \sum_{k_z=0}^{N_z-1} \hat{q} (\mathbf{k}, t) \exp(i \mathbf{k} \cdot \mathbf{x}) 
\end{equation}

where $\mathbf{k} =[k_x, k_y, k_z]^{\mathbf{T}}$ is the wavevector, $i=\sqrt{-1}$ and $\widehat{q}$ are the discrete complex Fourier coefficients of $q$. The Fourier coefficients can be computed with the fast Fourier transform (FFT) using the P3DFFT library. The main motivation of using a Fourier discretization is the simplicity of inverting a diagonal Laplace operator and the high numerical precision reflected in the absence of numerical diffusion and dissipation in the discretization. The gradient of a scalar can, for instance, be obtained by multiplying with the wavevector and the complex unit, $\widehat{\nabla q} = i \mathbf{k} \hat{q}$. The Laplace operator becomes a simple multiplication by $- |\mathbf{k}|^2$, it is thus diagonal in Fourier space. For further details, we refer the reader to the reference article on the FLUSI solver \cite{Flusi}.

\subsection{Fluid-structure interaction}

\begin{figure}[ht!]
\centering
\includegraphics[width=.9\linewidth]{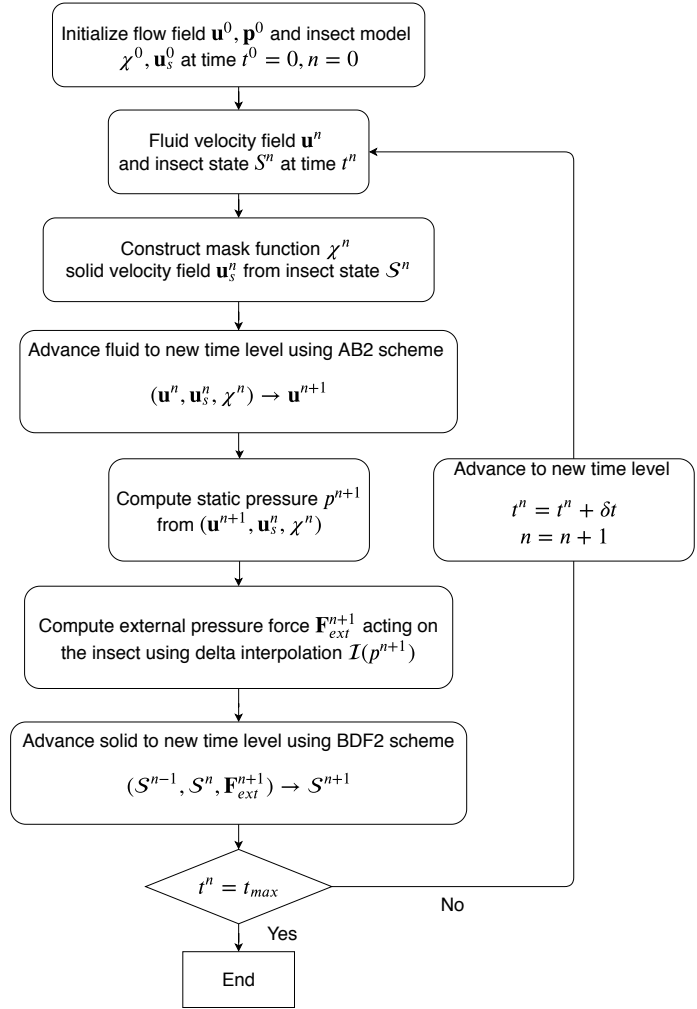}\\
  \caption{Flowchart: Semi-implicit staggered scheme of the time advancement for the fluid-structure interaction problem.}
  \label{fig:FSI_staggered_scheme}
\end{figure}

For time-stepping, the coupled fluid-solid system is advanced by employing a semi-implicit staggered scheme, as proposed in~\cite{ThomasThesis}. On the one hand, we advance the fluid by using the Adam--Bashforth second order (AB2) scheme with exact integration of the viscous term. On the other hand, the Backward Differentiation Formula of second order (BDF2) is used for the time discretization of the solid solver. The two modules are then coupled by the algorithm presented in the flowchart shown in figure~\ref{fig:FSI_staggered_scheme}. For the range of Reynolds numbers (75-4000), Dickinson et al.~\cite{DickinsonViscous1,DickinsonViscous2,PreidikmanViscous3} showed that pressure forces dominate the shear viscous forces. Hence, for calculating the solid deformation, the viscous fluid tension is considered negligible compared to the static pressure. Moreover, the scheme is called a weak coupling method since the static pressure is computed from the previous state of the solid model. This makes the system conditionally stable only if the structure is heavy enough with respect to the fluid density. However, the scheme is efficient because the fluid and the solid need to be advanced only one time at the current time level. Full details of the fluid–structure interaction (FSI) framework as well as detailed validation of the results can be found in our previous work~\cite{HungRevol}. 

\section{ Numerical set-up and bumblebee model }
\label{sec:setupbb}

To study the influence of wing flexibility on the aerodynamic forces, we compare the flexible wings with rigid ones using the same numerical set up in previous work in \cite{EKSLS16}.

\subsection{Flow configuration}

\begin{figure}[ht]
\centering
\includegraphics[width=1\linewidth]{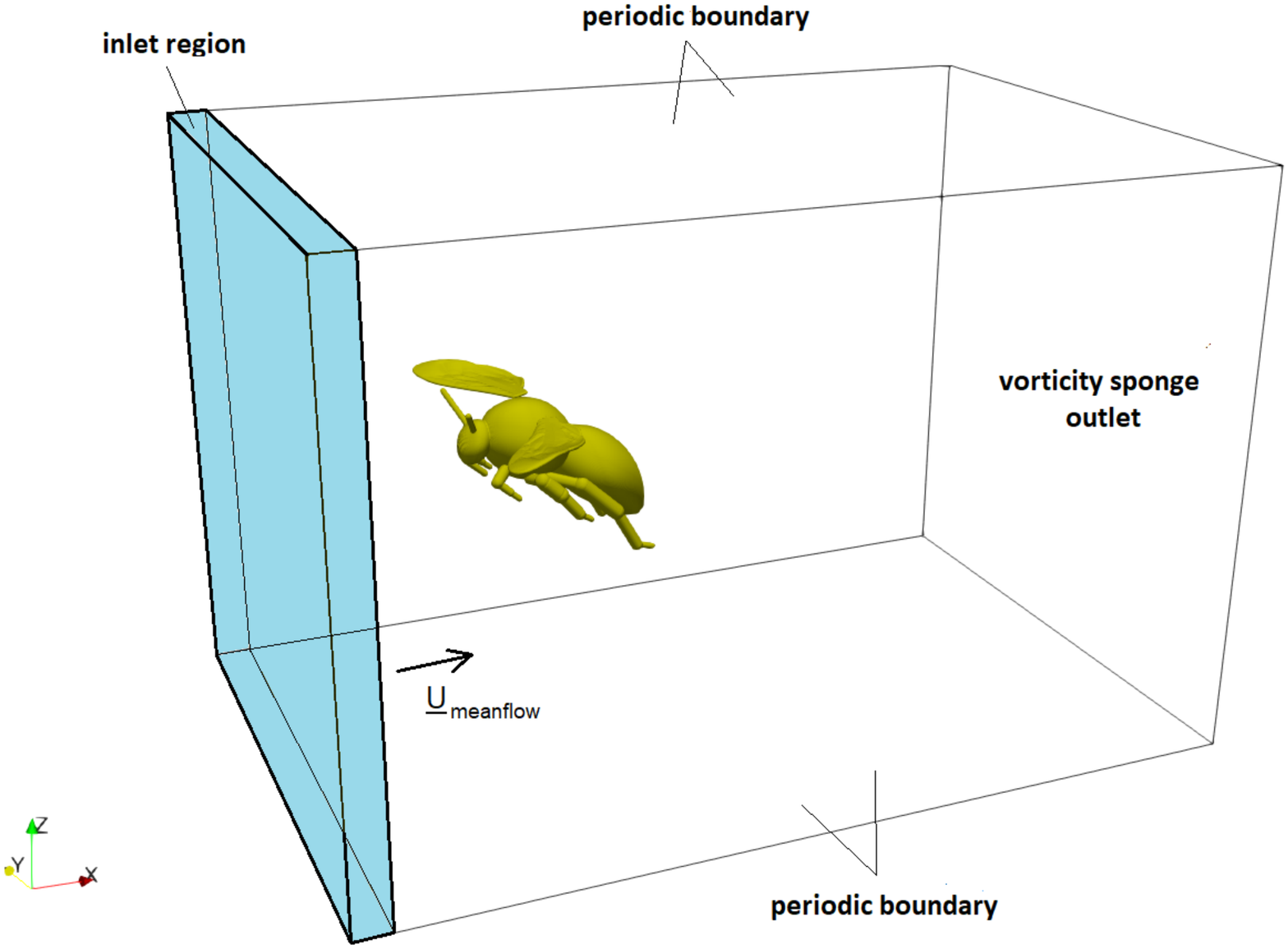}\\
  \caption{Illustration of the computational domain of size $6R \times 4R \times 4R$ used in all simulations. A bumblebee model with flexible wings is tethered at $\mathbf{x}_{cntr} = (2R,2R,2R)^T$ in a flow with the mean flow velocity $\mathbf{u}_{\infty} = (1.246Rf,0,0)^T$. The blue inlet region is used to impose a precomputed HIT velocity field which is constantly advected downstream by the mean flow. Another vorticity sponge region is placed at the outlet to damp out vortices. Periodic boundary conditions are set for the four other sides of the domain.}
  \label{fig:computation_domain}
\end{figure}

The computational domain, shown in figure~\ref{fig:computation_domain}, is $6R \times 4R \times 4R$ large, where $R$ is the bumblebee wing length, discretized by $1152 \times 768 \times 768$ grid points. The bumblebee is tethered (both translational and rotational motion of the body are inhibited) at $\mathbf{x}_{\mathrm{cntr}} = (2R,2R,2R)^T$ and exposed to a head wind with a mean flow accounting for the insect's forward velocity $\mathbf{u}_{\infty} = (1.246Rf,0,0)^T$, where $f$ is the wingbeat frequency. Due to the periodicity inherent to the spectral method, a thin vorticity sponge outlet, covering the last 4 grid points in $x$-direction, is used to minimize the upstream influence of the computational domain. The sponge penalization parameter $C_{sp}$ is usually set to a value larger than the permeability $C_{\eta}$, normally $C_{sp}=10^{-1}$. By construction, the sponge term is divergence-free to avoid the influence on the pressure field, which in turn would be modified even in regions far away from the sponge due to its nonlocality. A detailed discussion on the influence of the vorticity sponge can be found in \cite{Flusi}.

In nature, insects do not always fly in a calm, quiescent environment. Instead, they face, most of the time, many kinds of aerial perturbations such as gusty wind, vortices or turbulent flow generated by surrounding obstacles. Taking this into account, both laminar and turbulent flows are investigated here to study the role of wing flexibility under these two circumstances. For the laminar case, in the entire computational domain a mean flow $\mathbf{u}_{\infty}$ is imposed by simply setting the zeroth Fourier mode of the velocity $\mathbf{u}$~\cite{ThomasThesis}. On the other hand, information on turbulent flow conditions, which are experienced by flying insects in nature, remains an open question with limited data~\cite{CombesHIT}. However, for indoor wind tunnel experiments, isotropic or near-isotropic turbulence generated by a grid has been used as inflow condition to study the impact of turbulence on insect flight performance. Consequently, a homogeneous isotropic turbulence (HIT) is chosen as turbulent inflow in our present work in order to compare with the results obtained for rigid wings in~\cite{EKSFLS19}. For this purpose, in the inlet region containing the first 48 grid points along the axial direction, a precomputed HIT velocity field $\mathbf{u'}$ is added into the mean flow as velocity fluctuations $\mathbf{u}_{in} =  \mathbf{u}_{\infty} +  \mathbf{u'}$. The HIT field is then transported downstream by the mean flow and evolves dynamically like grid turbulence. In order to compare with the results from~\cite{EKSLS16}, we use here a HIT field characterized by the same parameters which are the turbulent intensity $Tu=u'_{RMS}/u_{\infty}=0.33$, the integral length scale $\Lambda=0.77R$ and the turbulent Reynolds number $Re_{\lambda}=u_{RMS} \lambda/ \nu = 129$, based on the Taylor-micro scale $\lambda=0.18$. More technical details on this approach can be found in~\cite{EKSLS16,EKSFLS19,ThomasThesis}.

\subsection{Bumblebee model}

The bumblebee model here is the same as the one used in~\cite{EKSFLS19} and derived from case BB01 in~\cite{DudleyBBWngKinematics}, except for the wings which will be introduced later in section \ref{subsec:Flexible_wing_model}.

The animal’s body mass, $M$, is $175~mg$, the gravitational acceleration $g=9.81 m/s^2$ and wing length $R$, amounts to $15~mm$. The bumblebee is composed of linked rigid bodies including the head, the thorax, the abdomen, all legs, the proboscis and the antennae.  These parts are circular elliptical or cylindrical sections joined by spheres, and the bilateral symmetry of the insect is assumed. The Reynolds number is $Re = U_{tip} c_m/ \nu_{air} = 2685$, where $U_{tip} = 2 \Phi R f = 9.15~m/s$ is the mean wingtip velocity, $c_m = 4.6~mm$ the mean chord length, $\nu_{air} = 1.568 \cdot 10^{-5}~m^2/s$ is the kinematic viscosity of air, $f = 152~Hz \, (T = 1/f = 6.6~ms)$ is the wingbeat frequency ($T$ is duration) and $\phi = 115^{\circ}$ is the wingbeat amplitude. The wingbeat kinematics are prescribed based on the work of Dudley and Ellington~\cite{DudleyBBWngKinematics}.

\subsection{Flexible wing model}
\label{subsec:Flexible_wing_model}

The two flexible wings of the insect are modeled using the mass-spring system as detailed in~\cite{HungRevol}. In the following we describe the venation pattern, the mass distribution and the flexural rigidity of the veins. 

\subsubsection{Venation pattern}

\begin{figure}[ht!]
\centering
\includegraphics[width=.9\linewidth]{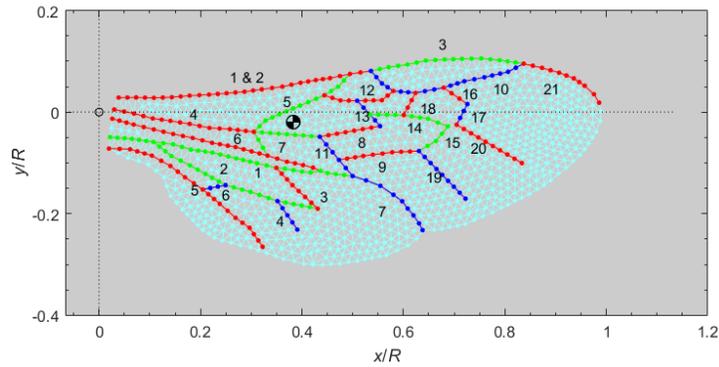}\\
  \caption{Illustration of the mass-spring model which is meshed based on measured data of real bumblebee wings~\cite{BumblebeeWingStructure}. The black and white marker represents the mass center. Color codes (red, green and blue) are used for identifying the veins and the membranes are represented by cyan circles.}
  \label{fig:wing_mesh}
\end{figure}

The venation architecture is claimed to be responsible for the anisotropy of the wing and it plays a crucial role on the wing dynamics during flight. Consequently, the functional approach is used to take into account the venation pattern in our model. The wing contour and the vein network are adapted from \cite{BumblebeeWingStructure} and encoded into the mass-spring system. The wing is then discretized by a triangular mesh with $1065$ mass points, as shown in figure~\ref{fig:wing_mesh}, using SALOME \footnote{https://www.salome-platform.org/}, an open-source integration platform for mesh generation. A mesh convergence study comparing between two wings, discretized by $465$ and $1065$ mass points, was performed in~\cite{HungRevol} for the revolving motion. Looking at the aerodynamic forces generated, the coarse-mesh wing showed no major difference with respect to the fine-mesh wing. However, for the flapping motion, the pressure field is expected to be more unstable and a fine-mesh wing is needed for the pressure interpolation in this case.

\subsubsection{Mass distribution}

The mass distribution represents the inertia of the system and the position of the mass center has a connection with the wing dynamics during flight. The mass distribution is calculated based on the measured wing mass data from \cite{BumblebeeWingStructure} and the vein pattern. For our numerical simulations, the total wing mass is chosen as the same used by Kolomenskiy et al.~\cite{BumblebeeWingStructure}, $m_w = 0.791~mg$. The mass is then distributed into vein and membrane parts based on their geometry and material. 

For the vein structure, each vein is considered as a rod composed of cuticle, $\rho_c = 1300~kg/m^3$ \cite{BumblebeeWingStructure}, with a circular cross section of constant diameter $d_v$ \cite{BumblebeeWingStructure} and length $l_v$, calculated directly from the model. The mass of each vein is then calculated and shown in~table~\ref{table:vein_mass}. Both diameter and mass are dimensionless quantities, normalized by wing length $R$ and air density $\rho_{air} R^3$, respectively. 
\begin{table}[]
\centering
\begin{tabular}{|c|c|c|ccc}
\hline
\multicolumn{3}{|c|}{Forewing}                                                                                           & \multicolumn{3}{c|}{Hindwing}                                                                                                                                                           \\ \hline
\# & \begin{tabular}[c]{@{}c@{}}Nominal\\ diameter\end{tabular} & \begin{tabular}[c]{@{}c@{}}Nominal\\ mass\end{tabular} & \multicolumn{1}{c|}{\#} & \multicolumn{1}{c|}{\begin{tabular}[c]{@{}c@{}}Nominal\\ diameter\end{tabular}} & \multicolumn{1}{c|}{\begin{tabular}[c]{@{}c@{}}Nominal\\ mass\end{tabular}} \\ \hline
1  & 0.0070                                                     & 0.0209                                                 & \multicolumn{1}{c|}{1}  & \multicolumn{1}{c|}{0.0065}                                                     & \multicolumn{1}{c|}{0.0180}                                                 \\ \hline
2  & 0.0074                                                     & 0.0237                                                 & \multicolumn{1}{c|}{2}  & \multicolumn{1}{c|}{0.0043}                                                     & \multicolumn{1}{c|}{0.0071}                                                 \\ \hline
3  & 0.0055                                                     & 0.0076                                                 & \multicolumn{1}{c|}{3}  & \multicolumn{1}{c|}{0.0046}                                                     & \multicolumn{1}{c|}{0.0024}                                                 \\ \hline
4  & 0.0070                                                     & 0.0063                                                 & \multicolumn{1}{c|}{4}  & \multicolumn{1}{c|}{0.0011}                                                     & \multicolumn{1}{c|}{0.0001}                                                 \\ \hline
5  & 0.0040                                                     & 0.0031                                                 & \multicolumn{1}{c|}{5}  & \multicolumn{1}{c|}{0.0038}                                                     & \multicolumn{1}{c|}{0.0043}                                                 \\ \hline
6  & 0.0048                                                     & 0.0094                                                 & \multicolumn{1}{c|}{6}  & \multicolumn{1}{c|}{0.0037}                                                     & \multicolumn{1}{c|}{0.0005}                                                 \\ \hline
7  & 0.0040                                                     & 0.0019                                                 & \multicolumn{1}{c|}{7}  & \multicolumn{1}{c|}{0.0020}                                                     & \multicolumn{1}{c|}{0.0012}                                                 \\ \hline
8  & 0.0038                                                     & 0.0009                                                 &                         &                                                                                 &                                                                             \\ \cline{1-3}
9  & 0.0041                                                     & 0.0023                                                 &                         &                                                                                 &                                                                             \\ \cline{1-3}
10 & 0.0048                                                     & 0.0064                                                 &                         &                                                                                 &                                                                             \\ \cline{1-3}
11 & 0.0045                                                     & 0.0017                                                 &                         &                                                                                 &                                                                             \\ \cline{1-3}
12 & 0.0038                                                     & 0.0018                                                 &                         &                                                                                 &                                                                             \\ \cline{1-3}
13 & 0.0042                                                     & 0.0010                                                 &                         &                                                                                 &                                                                             \\ \cline{1-3}
14 & 0.0038                                                     & 0.0020                                                 &                         &                                                                                 &                                                                             \\ \cline{1-3}
15 & 0.0034                                                     & 0.0008                                                 &                         &                                                                                 &                                                                             \\ \cline{1-3}
16 & 0.0032                                                     & 0.0005                                                 &                         &                                                                                 &                                                                             \\ \cline{1-3}
17 & 0.0032                                                     & 0.0004                                                 &                         &                                                                                 &                                                                             \\ \cline{1-3}
18 & 0.0044                                                     & 0.0009                                                 &                         &                                                                                 &                                                                             \\ \cline{1-3}
19 & 0.0015                                                     & 0.0001                                                 &                         &                                                                                 &                                                                             \\ \cline{1-3}
20 & 0.0018                                                     & 0.0001                                                 &                         &                                                                                 &                                                                             \\ \cline{1-3}
21 & 0.0020                                                     & 0.0009                                                 &                         &                                                                                 &                                                                             \\ \cline{1-3}
\end{tabular}

\caption{Dimensionless vein diameter $d_v$ (adapted from \cite{BumblebeeWingStructure}) and their corresponding dimensionless mass $m_v$.}
\label{table:vein_mass}
\end{table}

For the mass distribution of the membrane, the same optimization method as in \cite{HungRevol} is applied where the objective function is the difference between the mass center of the wing measured in the experiment~\cite{BumblebeeWingStructure} and the one calculated from the mass-spring model. For a mass point $m_i$ belonging to the membrane at position $[x_i, y_i]$, we get:
\begin{equation}
    m_i = 9.14 \cdot 10^{-5} - 3.48 \cdot 10^{-5} x_i + 2.48 \cdot 10^{-4} y_i
\end{equation}
Differences, between two mass centers, of $3.85 \cdot 10^{-3} [R]$ in the $x$-direction and $0.93 \cdot 10^{-3} [R]$ in the $y$-direction are obtained. These are negligible compared to the reference wing length $R$.

\subsubsection{Flexural rigidity of veins}

Because the bending stiffness of the membrane is neglected, the flexural rigidity of the wing comes solely from the flexural rigidity $EI$ of veins which is calculated based on their material and geometry. While the estimation of their second moments of inertia $I$ is straight forward using the diameter data from table~\ref{table:vein_mass}, determining the Young's modulus is not trivial. In our present work, the veins are considered to be made of cuticle which is reported to have a Young's modulus in the range of $1 kPa$ to $50 MPa$ \cite{CuticleProperties}. The wing needs to be flexible enough to reveal the influence of wing flexibility to the aerodynamic performance of insects but it cannot be too flexible to show unrealistic mechanical behaviors. For the purpose of our study, the value $E = 700~kPa$ is chosen.

\section{ Results and discussion }
\label{sec:results}

The forces generated by the bumblebee model with flexible wings as well as the required aerodynamic power will be presented in this section. Furthermore, they will be compared with the results obtained in \cite{EKSLS16} where the same bumblebee with rigid wings was considered. This allows us to have some insight into the wing flexibility influence on the insect aerodynamic performance.

\subsection{Tethered flight in laminar flow}

\begin{figure}[ht!]
\centering
\includegraphics[width=11.5cm]{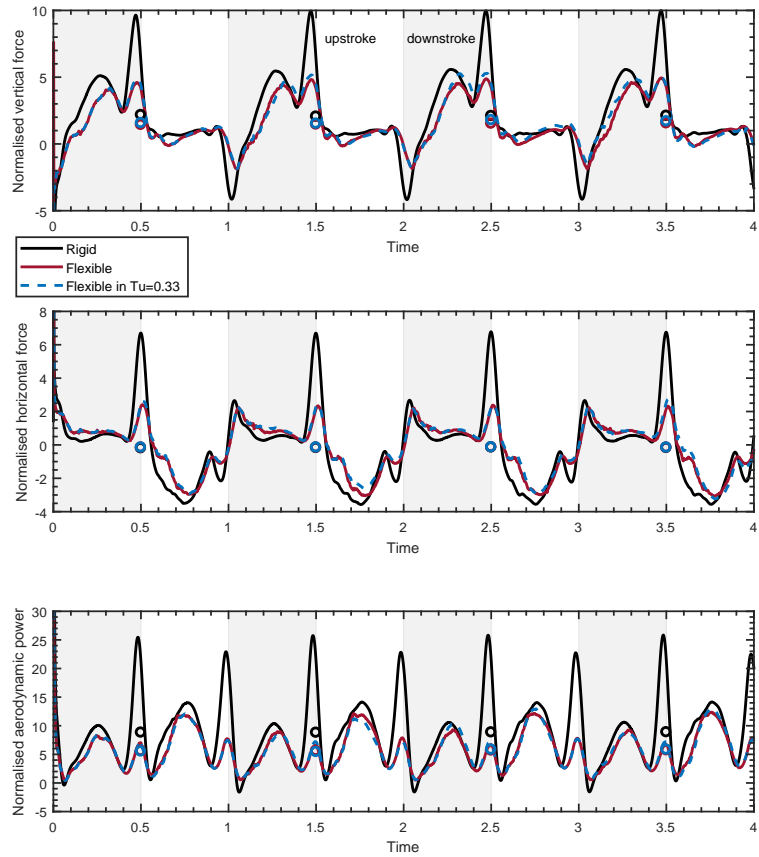}\\
  \caption{Normalized vertical force, horizontal force and aerodynamic power generated by a bumblebee with rigid wings (black) \cite{EKSFLS19} and flexible wings in laminar flow (red) and turbulent flow (blue). Circles represent the cycle-averaged value of forces and power.}
  \label{fig:forces_and_power}
\end{figure}

The vertical and horizontal forces produced by the flapping motion of the flexible wings are shown by red curves in figure~\ref{fig:forces_and_power} (a,b) while blue curves are those generated by rigid wings. The forces are normalised by $F = \rho_{air} R^4 f^2$. Here, the sideways force is small and not presented, since the animal is modeled with the assumption of symmetry. The simulation is computed for 4 strokes with 28776 time steps using 32 processors on Intel Xeon Gold 6142 (Sky Lake) $2.6 GHz$ and consumed 8128 CPU hours. For each cycle, the cycle-average values are calculated and presented in table~\ref{table:forces_and_power}. While the wing flexibility has minor effect on the average thrust with a decline of $11 \%$, it accounts for a $28 \%$ drop of the average lift. These losses can be explained as a result from the decrease of the effective angle of attack caused by wing deformation. The shape adaptation of the wing during the flapping motion alters the instantaneous angle of attack which is claimed to play a significant role in the force generation \cite{WShyyFlexAeroPeform}. However, these negative impacts do not necessarily mean that the rigid wings outperform aerodynamically their flexible counterparts. Although the flexible wings generate smaller forces, they consume much less energy, with almost $36 \%$ required aerodynamic power is reduced. The cycle-averaged lift-to-power ratio of flexible wing is $0.026$, $35 \%$ larger than the one of rigid wing which is $0.019$. 

Nevertheless, regarding the time evolution of the forces during one wingbeat, the instant surges of the forces at the ends of upstroke and downstroke, observed in the rigid case, are significantly weakened. The sudden rotation of the rigid wings at the midstrokes and the end of strokes are the reason for these large force peaks~\cite{SaneAeroInsectFlight}. This effect has now little impact due to the fact that the wing inertia are now taken into account. The inertial force makes the wing deform and streamline its shape to the airflow. This shape adaptation helps to mitigate the large pressure jump between upper and lower surfaces, especially at the trailing edge~\cite{WShyyLiftFlex} and provides a smoother flight~\cite{IfjuFlexMAV1,IfjuFlexMAV2}. This finding has more advantages in term of stabilizing generated forces, rather than lift-enhancement effect.

\begin{table}
\centering
\begin{tabular}{c c c c c c c}
\hline
Flow & \multicolumn{2}{c}{Thrust} & \multicolumn{2}{c}{Lift} & \multicolumn{2}{c}{Aerodynamic power} \\ [0.5ex] \ \
& Rigid & Flexible & Rigid & Flexible & Rigid & Flexible \\ [0.5ex] \hline \hline \\ [-1.5ex]
Laminar & 0.17 & 0.15 &  2.09 & 1.51 & 8.84 & 5.67  \\ [0.5ex] \hline
\end{tabular}
\caption{Cycle-averaged forces and power in the laminar case.}
\label{table:forces_and_power}
\end{table}

\subsection{Tethered flight in turbulent flow}

\begin{figure}[ht!]
\centering
\includegraphics[width=1\linewidth]{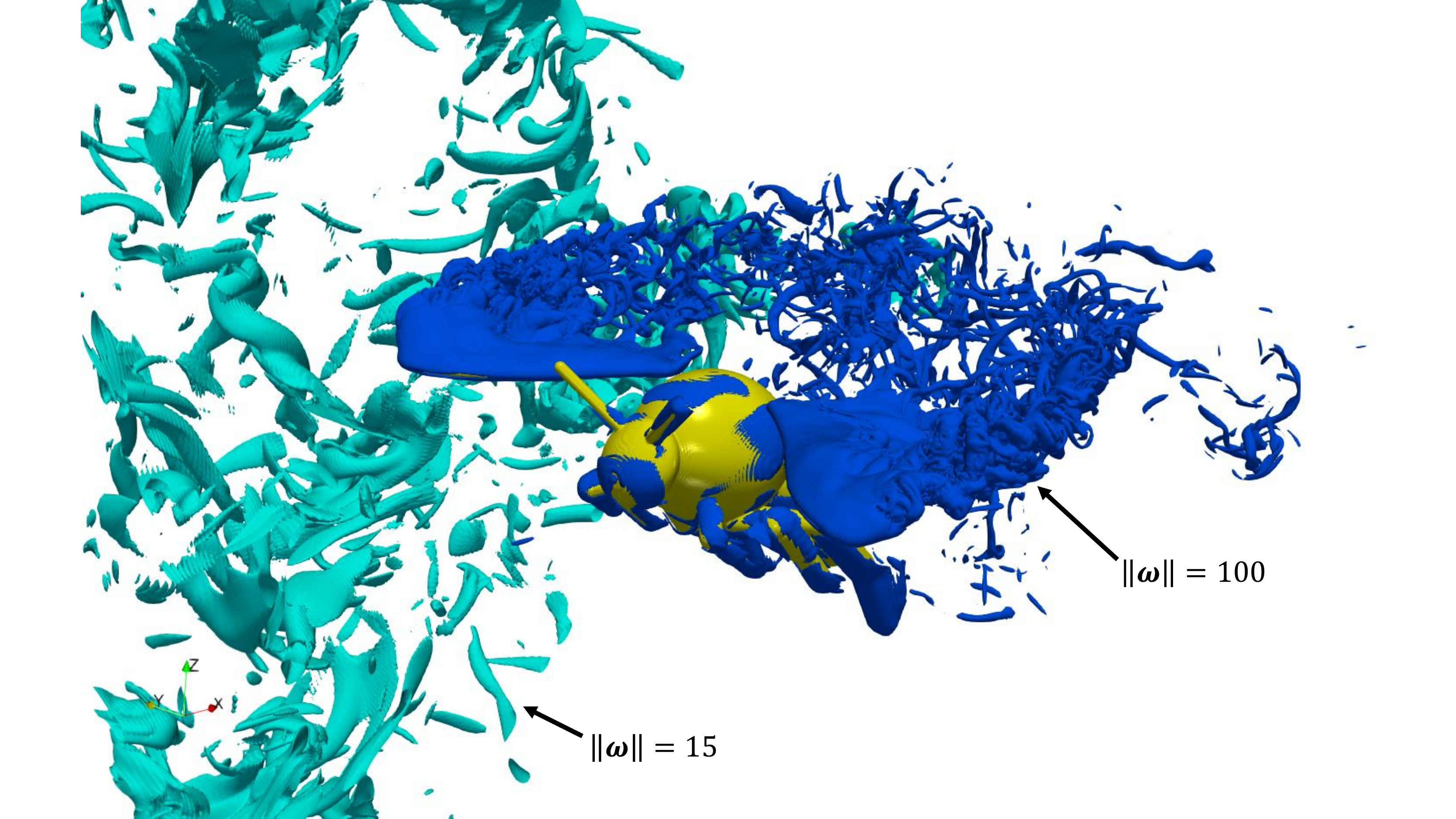}\\
  \caption{Visualization of flow generated by a tethered flapping bumblebee with flexible wings in turbulence $Tu=0.33$ showing normalized absolute vorticity isosurfaces at two levels $||\text{\boldmath$\omega$} || = 15$ (light blue) and $|| \text{\boldmath$\omega$} || = 100$ (dark blue). The flow fields are plotted at time $t=0.45/T$ and the weaker vortices are only shown in the region $3.7R \leq y \leq 4R$. }
  \label{fig:BB_in_turbulence}
\end{figure}

\begin{figure}[ht!]
\centering
\includegraphics[width=1\linewidth]{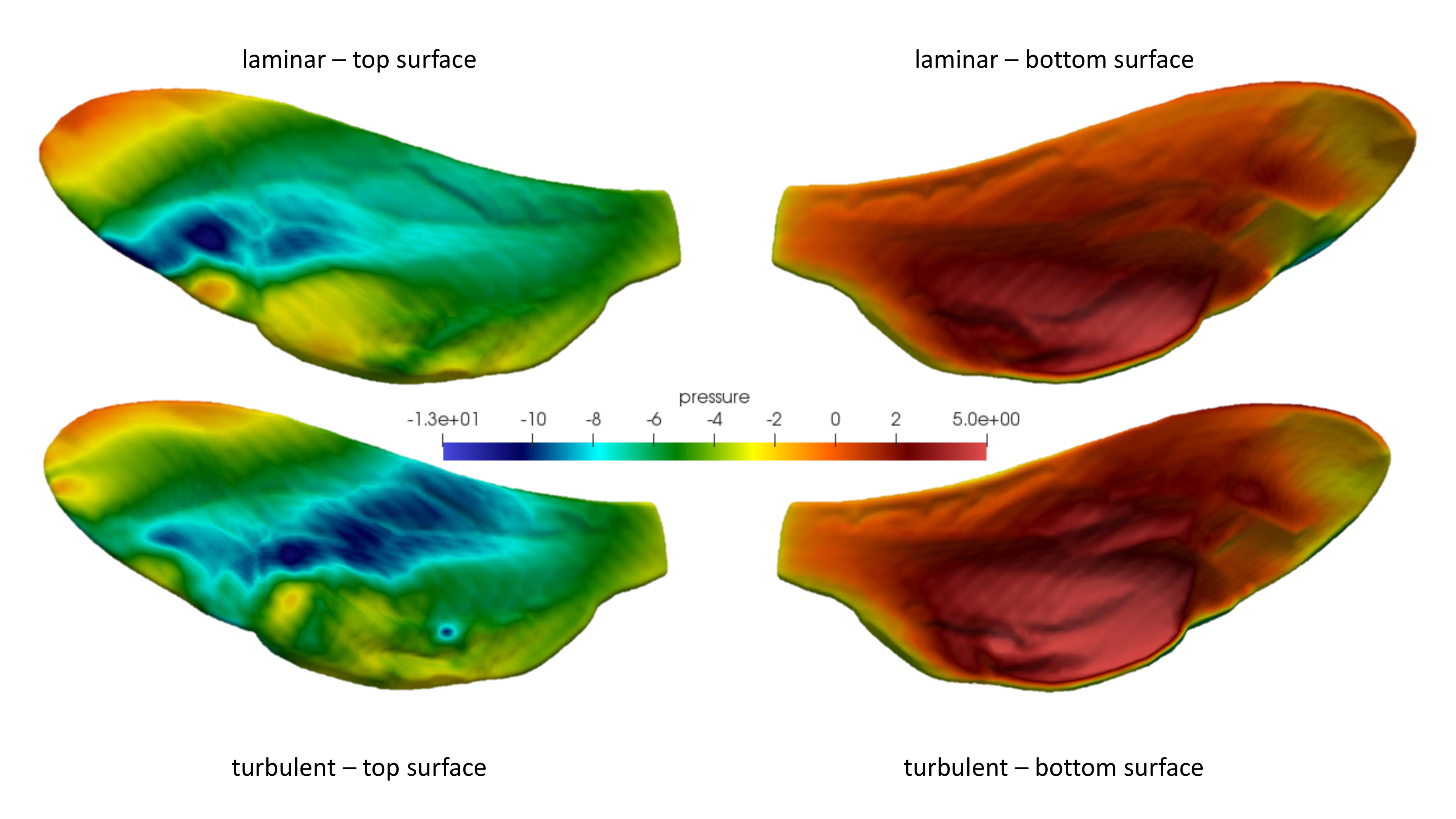}\\
  \caption{Normalized pressure distribution on the wing for top and bottom surfaces, plotted at time $t=0.45/T$ just before the stroke reversal, for the laminar (top) and turbulent cases (bottom).}
  \label{fig:comparison_pressure_distribution_on_wings}
\end{figure}

We then study the influence of an isotropic turbulence on the aerodynamic performance of a bumblebee by putting it in a turbulent flow. The simulation is computed for 4 strokes with 29000 time steps using 32 processors on Intel Xeon Gold 6142 (Sky Lake) $2.6 GHz$ and consumed 9000 CPU hours. Figure~\ref{fig:BB_in_turbulence} presents the flow structure of the bumblebee flying in a turbulent flow visualized by the normalized vorticity isosurfaces at two levels $\lVert \text{\boldmath$\omega$} \rVert = 15$ and $\lVert \text{\boldmath$\omega$} \rVert = 100$. The aerodynamic forces and the corresponding power in this turbulent condition are shown in figure~\ref{fig:forces_and_power}. The results demonstrate insignificant differences between turbulent and laminar flow conditions.  The aerodynamic forces generated by the bumblebee are almost identical to those derived during unperturbed, laminar inflow, with the same required energetic cost. For $Re>100$, the aerodynamic forces are mainly produced by the differential dynamics pressure across the wing \cite{SaneAeroInsectFlight}. Figure~\ref{fig:comparison_pressure_distribution_on_wings} shows the normalized pressure distribution on top and bottom wing surfaces of the two cases just before the stroke reversal $t=0.45~T$. The effect of turbulence can hardly be seen here which explains the negligible change of aerodynamic forces. The outcome here is consistent with the one observed in the rigid case in \cite{EKSLS16}. 

\section{ Conclusions and perspectives }
\label{sec:conclusions}

Following our previous work on revolving flexible wings~\cite{HungRevol}, the impact of wing flexibility was now studied in the context of tethered flight using flapping wing kinematics measured in experiments by Dudley and Ellington~\cite{DudleyBBWngKinematics}. High-resolution numerical simulations on massively parallel machines were carried out to solve the fluid-structure interaction problem between the fluid solver FLUSI and the solid solver based on a mass-spring system. Both laminar and turbulent inflows were considered to investigate diverse flight conditions of insects. The preliminary results obtained in this work allow us to have some understanding about the role of wing flexibility in flapping flight. 

In laminar flow, the aerodynamic forces and the required power have been calculated and compared with the ones obtained for rigid wings. We found that wing flexibility hardly contributed to lift or thrust enhancement. However, the significant reduction of the required power suggested that wing flexibility plays an important role in saving flight energetic cost. Moreover, the wing inertia also helped to damp out the fluctuation of the aerodynamic force and helped thus the insect to stabilize during flight.

In turbulent flow, although the ability of shape adaptation of flexible wings makes them more sensitive to fluctuation of the flow structure than their rigid counterparts, the impact of turbulence is still negligible under the considered flight conditions. Nevertheless, due to costly computational time, the statistical property of the turbulent flow is not considered because only one simulation is done to obtain the results for the turbulent case.nMoreover, due to the expensive computational cost, especially in the turbulent case, the mesh convergence study was not performed in this paper and we refer readers to \cite{HungRevol}.

Despite of these findings, we have to keep in mind that the wing kinematics has an essential effect on the aerodynamic performance of wings and we have considered only one set of wing motion in this study. In perspective, these limitations can be overcome by examining other species with different wing kinematics or including flight control in our model. This is planned for our work in the future where we will study Calliphora with its wing kinematics measured from experiments.  

Finally, although the wing flexibility was calculated based on the geometrical property of the veins, the estimation of veins' Young's modulus remains somewhat limited due to the vast range of known cuticle's property. This can be improved by using mathematical optimization for determining the right elastic properties of the wing model. To this end the equilibrium state of the wing model under external static force as a function of wing stiffness will be calculated and compared with data measured from experiments done by our team.
%

\section*{Acknowledgements} 
Financial support from the Agence Nationale de la Recherche (ANR Grant No. 15-CE40-0019) and Deutsche Forschungsgemeinschaft (DFG Grant No. SE 824/26-1), project AIFIT, is gratefully acknowledged.
The authors were granted access to the HPC resources of IDRIS under the Allocation No. 2018-91664 attributed by GENCI (Grand \'Equipement National de Calcul Intensif). For this work, Centre de Calcul Intensif d’Aix-Marseille is acknowledged for granting access to its high performance computing resources financed by the project Equip{@}Meso (No. ANR-10-EQPX- 29-01).
The authors thankfully acknowledge financial support granted by the minist\`eres des Affaires \'etrang\`eres et du d\'eveloppement international (MAEDI) et de l'Education nationale et l'enseignement sup\'erieur, de la recherche et de l'innovation (MENESRI), and the Deutscher Akademischer Austauschdienst (DAAD) within the French-German Procope project FIFIT.

D.K. gratefully acknowledges financial support from the JSPS KAKENHI Grant No. JP18K13693.

%
%
%

\end{document}